\documentclass[prl,aps,twocolumn,groupedaddress,showpacs]{revtex4}

\usepackage{pxfonts}
\usepackage{graphicx}
\usepackage{color}
\begin{document}

\title{Drift-diffusion model of the fragmentation of the external ring structure in the photoluminescence pattern of indirect excitons in coupled quantum wells}

\author{J. Wilkes}
\author{E.\,A. Muljarov}
\author{A.\,L. Ivanov$^\dagger$}
\affiliation{Department of Physics and Astronomy, Cardiff University, Cardiff CF24 3AA, United Kingdom}

\begin{abstract}
Under optical excitation, coupled quantum wells are known to reveal fascinating features in the photoluminescence pattern originating from dipole orientated indirect excitons. The appearance of an external ring has been attributed to macroscopic charge separation in the quantum well plane. We present a classical model of non-linear diffusion to account for the observed fragmentation of the external ring into a periodic array of islands. The model incorporates the Coulomb interactions between electrons, holes and indirect excitons. At low temperatures, these interactions lead to pattern formation similar to the experimentally observed ring fragmentation. The fragmentation is found to persist to temperatures above the quantum degeneracy temperature of indirect excitons.
\end{abstract}

\pacs{71.35.-y, 78.55.Cr, 89.75.Kd}

\date{\today}

\maketitle

Since the first theoretical predictions that excitons - bound states of electrons and holes - may undergo Bose Einstein condensation (BEC) \cite{KeldyshKozlov}, intensive studies have sought to realize the phenomenon experimentally. This arduous journey has inspired a diverse array of experiments and investigations \cite{SnokeNAT418,ButovNAT418,HammackPRB76,VorosPRL97,ButovNAT417}. Novel structures and techniques have been employed which, regardless of their success or failure, have deepened our understanding of the excitonic properties of semiconductors.

A system which remains a strong candidate for observing exciton BEC is that of indirect excitons in coupled quantum wells (CQWs) \cite{SnokeNAT418,ButovNAT418}. The samples consist of two thin layers of semiconductor material stacked between layers with a higher band gap energy. A voltage applied in the growth direction shifts the electron and hole energies with respect to each other. In this arrangement, optical excitations create indirect excitons where electrons and holes are confined to adjacent quantum wells by tunneling through the barrier region. The spatial separation of electron and hole wave functions extends the optical lifetimes of excitons well beyond the time needed for thermalization to the lattice temperature \cite{ButovPRB59,Sivalertporn,IvanovJPCM16,IvanovEPL73,WilkesPRB80}. These structures provide optimal conditions for creating a cold dense exciton gas - a prerequisite for BEC.

Some of the first results from this type of experiment, published nearly a decade ago \cite{SnokeNAT418,ButovNAT418}, revealed some remarkable features in the spatial photoluminescence (PL) pattern of indirect excitons. These include the inner ring and localized bright spots. Of interest here is the bright {\it external ring} which can be greater than ${\rm 100\,\mu m}$ in diameter and is separated from the central bright spot around the laser by a predominantly dark region. Previous studies \cite{RapaportPRL92,SnokeSSC127,ButovPRL92} have shown that the ring forms at the interface between spatially separated electron-rich and hole-rich regions. This is due to a difference in the capture efficiency of the injected electrons and holes which results from their different masses. An abundance of holes builds around the laser spot as some fraction of electrons leaks to the electrodes. In the presence of an electric field, the doping in the electrode layers leads to a background electron concentration within the wells. The result is a pool of holes surrounded by a sea of electrons with excitons forming at the boundary.

Perhaps the most remarkable feature observed is the fragmentation of the external ring into a periodic array of islands for temperatures below $\approx 3\,{\rm K}$ \cite{ButovNAT418,YangPRB75,YangPRL97}. We note that a previous model \cite{LevitovPRL94} explains the macroscopic ordering via a process of stimulated scattering of excitons into the ground state whilst another model \cite{ParaskevovPLA368} assumes a BEC present in the ring. Both imply that fragmentation is a signature of quantum degeneracy.

An alternative model explained the ring fragmentation in terms of attractive interactions between excitons \cite{SugakovSSC134,ChernyukPRB74}. In this framework, a ring of uniform density becomes unstable and the attractive interaction leads to droplet formation. However, measurements of the blue shift in the exciton emission \cite{YangPRB75} reveal clear evidence that the interaction is repulsive and that this mechanism cannot be responsible for the effect.

In this letter, we present a model to asses the importance of Coulomb interactions in the external ring. We find that at low temperatures, these interactions lead to periodic modulation of the exciton density. Whilst spatial coherence measurements suggest a statistically degenerate exciton gas is present in the ring \cite{YangPRL97,HighNAT483}, we illustrate that fragmentation can occur due to classical mechanisms and cannot unambiguously be taken as evidence of degeneracy. Studies of the role of Coulomb interactions in this system have been attempted previously \cite{ParaskevovPRB81,DenevSSC134} and have highlighted their effect on the appearance of the external ring. However, due to the computational complexity of the problem, these works have been limited to the case of 1D geometry or to a low density regime and are insufficient to capture fragmentation of the ring. In Ref.\,\cite{DenevSSC134}, it is shown that inclusion of Coulomb terms is necessary to explain the threshold dependence of the ring radius on excitation power and confirms the significance of the contribution made by the interactions included here. 

Conceptually, we find that at low temperatures, the diffusive nature of carrier transport diminishes. In this regime, carriers move under the action of strong electric fields provided mainly by the in-plane charge separation of electrons and holes. Under these conditions, a ring of uniform density is an unstable configuration and the system naturally evolves to a fragmented ring as a process of energy minimization.

Based on the approach used in \cite{RapaportPRL92,ButovPRL92,DenevSSC134,HaquePRE73}, the following set of coupled equations were used to model the creation, transport and decay of carriers in CQWs:
\begin{eqnarray}
\frac{\partial n_{\rm e}}{\partial t} &=& \nabla \left[D_{\rm e} \nabla n_{\rm e} + \mu_{\rm e} n_{\rm e} \nabla U_{\rm e} \right] + \frac{n_{\rm e}^{(0)} - n_{\rm e}}{\tau_{\rm e}} + \Lambda_{\rm e} - w n_{\rm e} n_{\rm h}, \nonumber \\
\frac{\partial n_{\rm h}}{\partial t} &=& \nabla \left[D_{\rm h} \nabla n_{\rm h} + \mu_{\rm h} n_{\rm h} \nabla U_{\rm h} \right] + \Lambda - w n_{\rm e} n_{\rm h}, \\
\frac{\partial n_{\rm x}}{\partial t} &=& \nabla \left[D_{\rm x} \nabla n_{\rm x} + \mu_{\rm x} n_{\rm x} \nabla U_{\rm x} \right] + w n_{\rm e} n_{\rm h} - \frac{n_{\rm x}}{\tau_{\rm x}}. \nonumber
\end{eqnarray}
The solution is the density distributions $n_i$ of electrons, holes and excitons ($i = \rm{e, h, x}$ respectively). The first and second terms inside each of the square brackets are the diffusion and drift currents respectively. $D_i$ and $\mu_i$ are the diffusion coefficients and mobilities. Each diffusion coefficient is assumed constant and we use the classical limit for the mobilities, $\mu_i \approx D_i/(k_B T)$. Far from the laser induced heating at the excitation spot, the temperature of each species is defined by the lattice temperature, $T$. Coulomb interactions are included via the potentials $U_{\rm e}$, $U_{\rm h}$ and $U_{\rm x}$. For electrons and holes residing in adjacent quantum wells, one has
\begin{eqnarray}
U_{\rm e} &=& n_{\rm e} * V_0 \,\,\, - \,\,\, n_{\rm h} * V_d \,\,\, + \,\,\, n_{\rm x} * (V_0 - V_d), \nonumber \\
U_{\rm h} &=& n_{\rm h} * V_0 \,\,\, - \,\,\, n_{\rm e} * V_d \,\,\, + \,\,\, n_{\rm x} * (V_0 - V_d), \\
U_{\rm x} &=& (n_{\rm e} + n_{\rm h} + 2 n_{\rm x}) * (V_0 - V_d), \nonumber
\end{eqnarray}
where $V_\alpha({\bf r}) = e^2/ (4 \pi \varepsilon_0 \varepsilon_r \sqrt{|{\bf r}|^2 + \alpha^2} )$ and '$*$' denotes the convolution; $f * g = \int f({\bf r} - {\bf r}') g({\bf r}') d {\bf r}'$. Here, ${\bf r}$ is the in-plane coordinate. All calculations are made for $\rm GaAs/Al_{0.33}Ga_{0.67}As$ CQWs studied in Ref.\cite{ButovNAT418}. We use $d=12\,{\rm nm}$ for the separation between electron and hole layers and dielectric constant $\varepsilon_r = 12.9$. Both $d$ and $\varepsilon_r$ are critical parameters in the onset of fragmentation since they scale the strength of the interaction terms. Macroscopic charge separation creates an in-plane potential gradient under which electrons and holes drift towards the ring position. The separation of opposite charges into adjacent wells leads to repulsion in the exciton-exciton interaction and also in the exciton-electron and exciton-hole interactions. The interactions make a significant contribution at low temperatures where transport due to Coulomb forces dominates over diffusive mechanisms.
\begin{figure}
\begin{center}
\includegraphics[width=220pt]{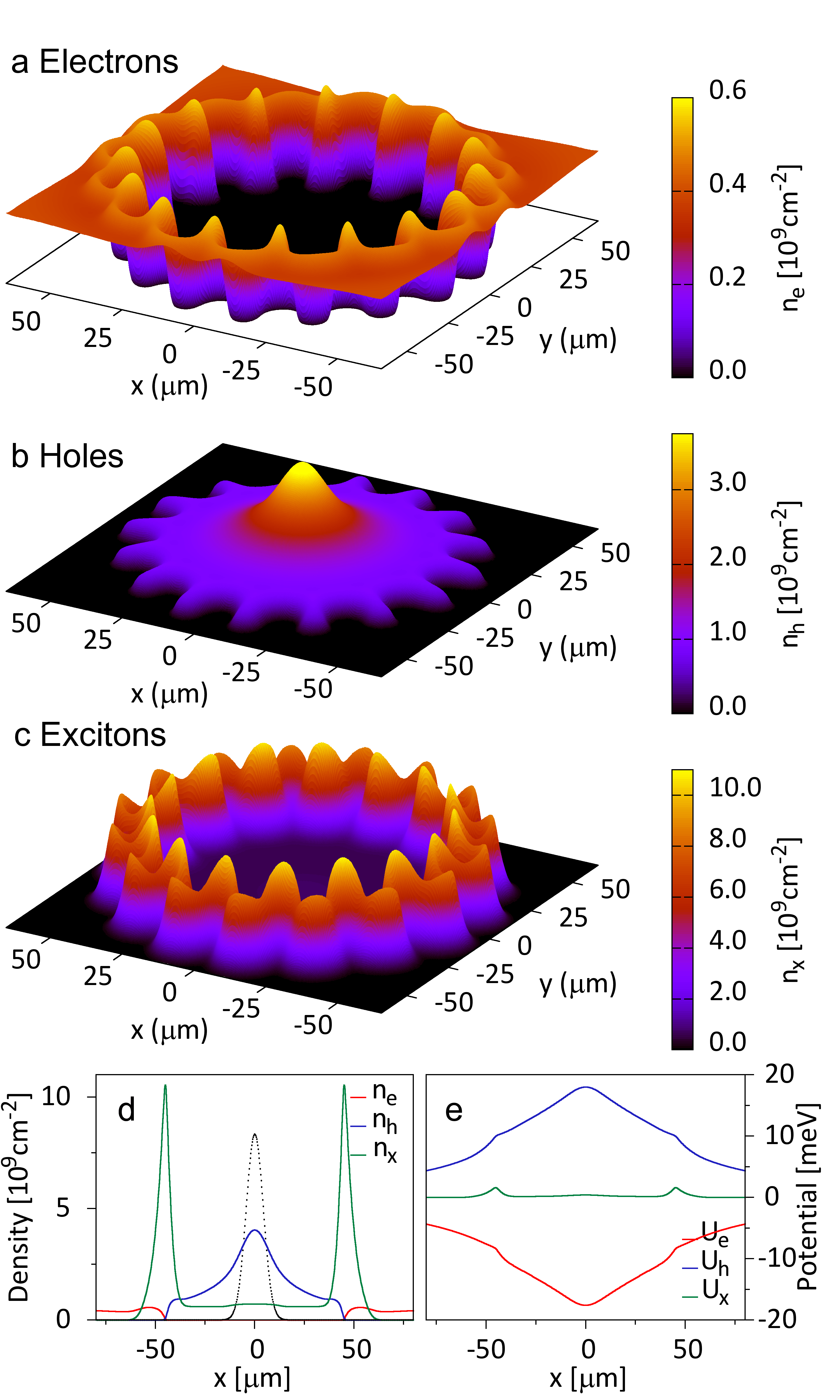}
\caption{Spatial density distributions for electrons (a), holes (b) and indirect excitons (c). An optical pump of FWHM $\rm 10\,\mu m$ is focused at the origin with peak generation rate of $\rm 2 \times 10^{10} cm^{-2} ns^{-1}$. Background electron density $n_{\rm e}^{(0)} = {\rm 0.5 \times 10^9 cm^{-2}}$. Other parameters are given in the main text. A cut along the $x$-axis of the density distributions (d) and potential (e). The pump profile is shown by black dots.}
\label{fig1}
\end{center}
\end{figure}

The remaining terms on the right hand side of Eq. (1) account for the creation and decay of each species. The homogeneous source term for electrons, which is due to a constant flux through the CQWs, acts to restore a background density, $n_{\rm e}^{(0)}$ in a characteristic time $\tau_{\rm e}$. Both $n_{\rm e}^{(0)}$ and $\tau_{\rm e}$ depend on the applied electric field. To reduce the number of control parameters, we modeled control of the electric field by varying $n_{\rm e}^{(0)}$ and fixed $\tau_{\rm e}=50\,{\rm ns}$ which is close to the experimentally determined value in \cite{YangPRB81}. $\Lambda({\bf r})$ is the injection rate of unbound electron-hole pairs into the CQWs by the laser. The imbalance in the generation rate of electrons and holes can be incorporated by $\Lambda_{\rm e} < \Lambda$. Without loss of generality, we set $\Lambda_{\rm e} = 0$ and compensate by reducing $\Lambda$. This removes the central bright spot from simulations. The binding rate of free pairs into excitons is proportional to the overlap of the electron and hole densities. The fit parameter $w$ is inversely proportional to the exciton formation time and fragmentation is sensitive to its value. Here we used $w=10^3{\rm cm^2s^{-1}}$\cite{PiermarocchiPRB55}. This leads to an exciton density of about $10^{10}{\rm cm^{-2}}$ which has been determined experimentally via the exciton blue shift \cite{ButovNAT418}. Reducing $w$, reduces $n_{\rm x}$ and diminishes the interactions, preventing ring fragmentation. This term is a decay channel for electrons and holes and a source term for excitons. $\tau_{\rm x}$ is the optical lifetime of excitons and is nearly constant with respect to $n_{\rm x}$ \cite{WilkesPRB80}. To simplify the model, we fixed $\tau_{\rm x}=50\,{\rm ns}$ and have not explored the effect of its dependence on temperature or electric field. $D_{\rm x}=0.2\,{\rm cm^2s^{-1}}$ was used for the exciton diffusion constant. This is consistent with the model used in \cite{WilkesPRB80} where the effect of a QW disorder potential of $\approx 1\,{\rm meV}$ was included. The electron and hole diffusion constants are $D_{\rm e}=30\,{\rm cm^2s^{-1}}$ and $D_{\rm h}=15\,{\rm cm^2s^{-1}}$, comparable to experimentally measured values in \cite{YangPRB81}.

Figs.\,1a-d show the density distributions satisfying (1-2). The equations were solved dynamically with the potentials $U_i$ being continuously recalculated on each time step until a steady state solution reached. The simulations use initial conditions of uniform density $n_{\rm e} = n_{\rm e }^{(0)}$ and $n_{\rm h} = n_{\rm x} = 0$. At a circular boundary well beyond the ring position, the electron density is fixed to $n_{\rm e}^{(0)}$ and the hole and exciton densities to zero. A Gaussian excitation profile for $\Lambda({\bf r})$ is focused on the center of the sample. The results show a periodic modulation of the exciton density along the ring. Additionally, a slight modulation of the ring radius is seen which is not observed in experiments. Figs. 1d and 1e show cross sections along the line $y=0$ for the density distributions and potentials respectively. In-plane charge separation causes an E-field of $\approx 2\,{\rm eV cm^{-1}}$ driving electrons and holes to the ring position thus enhancing the generation rate of excitons.

A feature of the experimental data which is well reproduced by the model is the dependence of the ring fragmentation on the lattice temperature. Fragmentation is observed only below a critical temperature \cite{ButovNAT418,YangPRB75}. This has been interpreted as the indirect exciton degeneracy temperature where the statistics cross from the classical to the quantum regime \cite{LevitovPRL94}. In our model, the lattice temperature $T$ appears in the denominator of the interaction terms which, as a result, become greater in magnitude than the diffusive terms for $\nabla U_i > k_{\rm B} T (\nabla n_i)/n_i$.
\begin{figure}
\includegraphics[width=245pt]{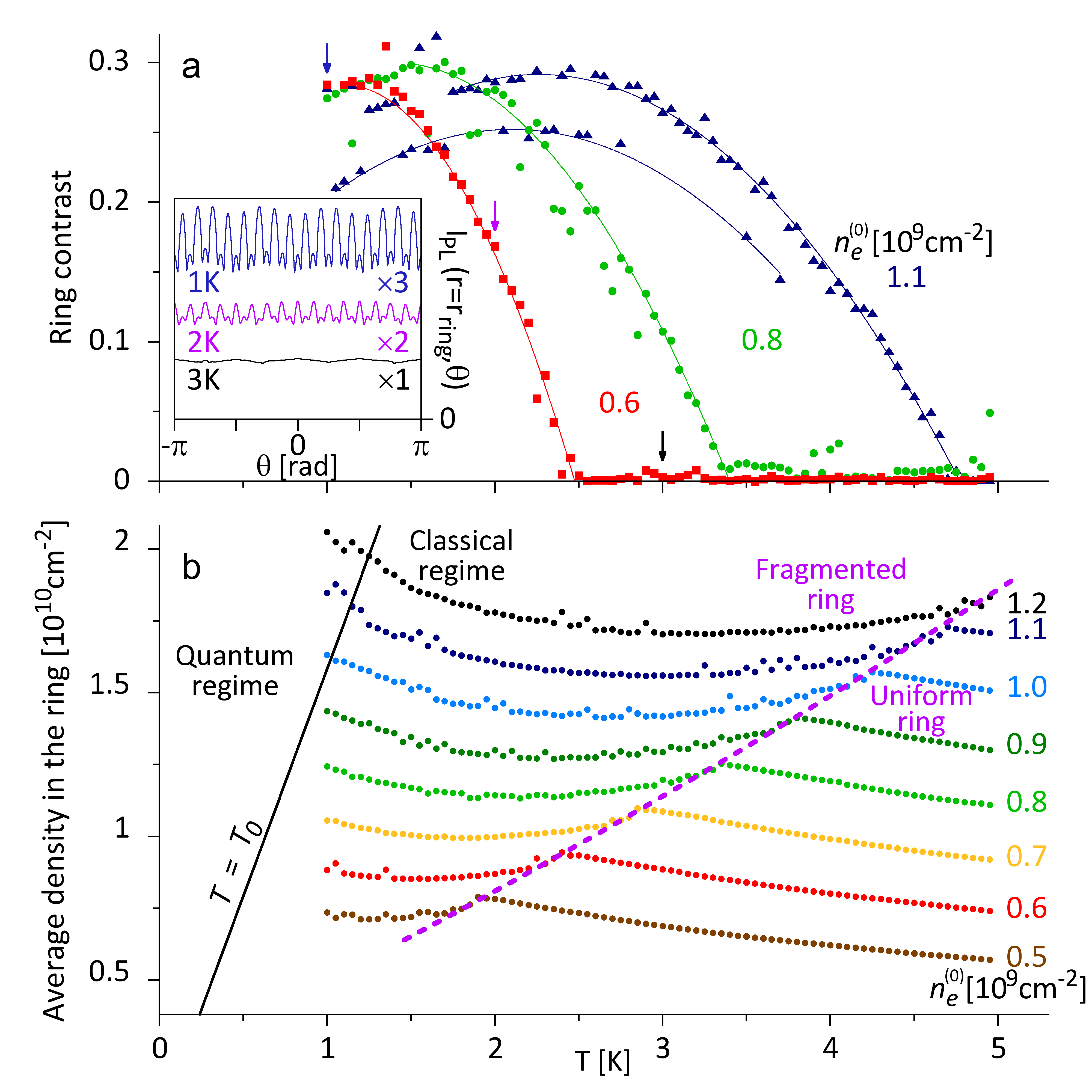}
\caption{a) Ring contrast, $\langle(I_{\rm PL}^{\rm max} - I_{\rm PL}^{\rm min})/I_{\rm PL}^{\rm max}\rangle$ against temperature for various $n_{\rm e}^{(0)}$. Average ring radius is kept constant by adjusting $\Lambda$. Inset: PL intensity profiles $I_{\rm PL}$ along the ring for the points on the red curve shown by arrows. b) Average density in the external ring against temperature. The dashed line marks the transition from a uniform to a fragmented ring. The solid line marks the crossing from classical to quantum statistics.}
\label{fig2}
\end{figure}

The inset in Fig\,2a shows the exciton density profile at the ring position, $r_{\rm ring}$ for various temperatures. The ring position is defined as the location of maximum exciton density for each angle $\theta$ about the excitation spot. Ring fragmentation appears with decreasing temperature as the drift currents dominate over the diffusive currents. This is illustrated in the main panel where the average contrast between the peaks and dips along the ring is plotted against $T$. The model captures the temperature dependence of the pattern formation observed in experiments \cite{ButovNAT418,YangPRB75}. The onset of fragmentation is abrupt and a critical temperature can be identified. This temperature is deceivingly low and could be misinterpreted as the degeneracy temperature. The critical temperature varies with exciton density which can be controlled by adjusting $\Lambda$ and $n_{\rm e}^{(0)}$ simultaneously whilst maintaining a fixed average radius. Each curve corresponds to a different value of $n_{\rm e}^{(0)}$ with $\Lambda$ chosen to give a radius of $50\,{\rm \mu m}$. The seemingly noisy data in Fig\,2a is a consequence of the state of the ring evolving to configurations with different numbers of islands. The degenerate solutions have slightly different contrasts and the effect is most pronounced at the highest density.

Fig\,2b shows the spatially averaged density in the ring against temperature for different values of $n_{\rm e}^{(0)}$. The dashed line marks the average density at each critical temperature. Also shown is the line ${T = T_{\rm 0}}$. Here, ${T_{\rm 0} = (\pi \hbar^2 n_{\rm x})/(2 M_{\rm x} k_{\rm B})}$ is the exciton degeneracy temperature ($M_{\rm x} = 0.22 m_0$ is the exciton mass). The model predicts that fragmentation can occur both well above and below ${T_{\rm 0}}$. At the highest densities examined, an increase in exciton density is observed as the temperature is further reduced beyond the onset of fragmentation. In PL experiments, this manifests itself as an increase in the blue shift in the exciton spectrum which has been observed in Ref.\,\cite{YangPRB75}. In that work, the blue shift was found to also increase with temperature when above the critical temperature. Our model would be consistent with this if the temperature dependent $\tau_{\rm x}$ described in \cite{IvanovEPL73} was used.

Due to doping, there is an abundance of free electrons in the layers surrounding the quantum wells. Arguably, these electrons may accumulate around the excitation spot to screen the in-plane potentials $U_{\rm e}$ and $U_{\rm h}$ which would suggest that Coulomb interactions are not prominent enough to be responsible for the fragmentation of the ring. However, on comparison of available experimental data, an alternative picture presents itself. Ring fragmentation is seen in the work of Butov {\it et al.}\cite{ButovNAT418}, but not in the work by Snoke {\it et al.}\cite{SnokeNAT418}, where it is estimated that a much greater density of free electrons is present in the layers adjacent to the CQWs. From this, we can deduce that in the former there are insufficient free electrons to screen the Coulomb interactions and prevent ring fragmentation. The detailed effect of a free electron gas in the adjacent layers is beyond the scope of this work and its effect on pattern formation is an open question.

To further understand the physical mechanism that leads to ring fragmentation, it is instructive to examine the case of a line excitation. In this geometry the laser is focused not to a single point but to a narrow line extended across the entire sample. This leads to the appearance of two parallel {\it external lines} in the indirect exciton PL pattern, either side of the laser.
\begin{figure}
\includegraphics[width=245pt]{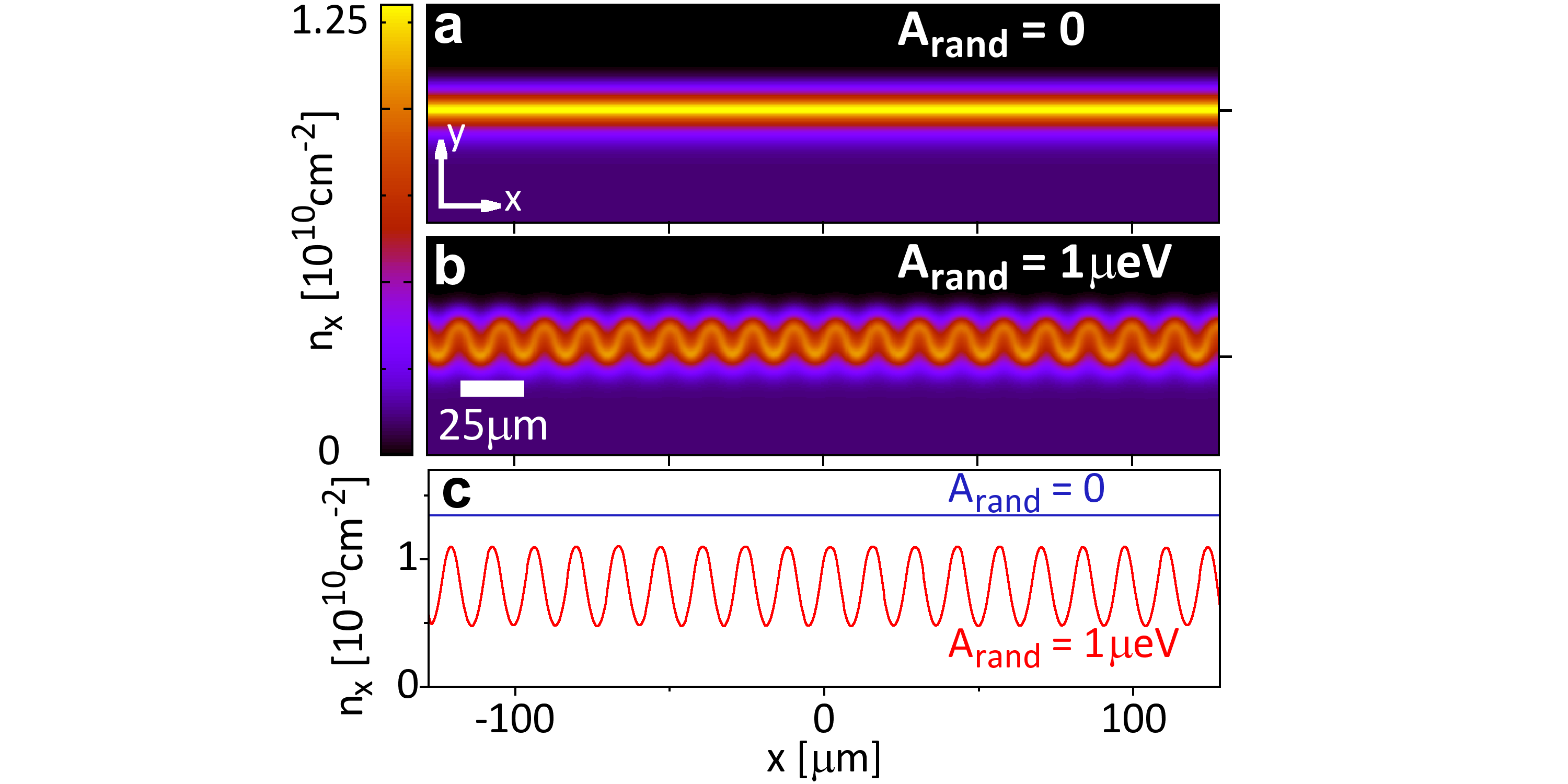}
\caption{Exciton density distribution due to a line excitation for $A_{\rm rand}=0$ (a) and $A_{\rm rand}=1\,{\rm \mu eV}$ (b) at $T = 1\,{\rm K}$ and $n_{\rm e}^{(0)} = 10^9\,{\rm cm^{-2}}$. The density along the external line with and without disorder at $T = 1\,{\rm K}$ (c). Periodic boundaries are at $y=\pm 128{\rm \mu m}$ and $n_{\rm e}^{(0)} = 10^9\,{\rm cm^{-2}}$.}
\label{fig3}
\end{figure}
In Fig. 3 we present results for this situation. The laser has a Gaussian profile in the $y$ direction and is homogeneous in the $x$ direction. Periodic boundary conditions are used to simulate an infinitely extended line. Unlike the external ring in the point excitation geometry, the external line does not spontaneously fragment. Instead, a small perturbation is needed to observe the effect. This is included by adding a random disorder potential to either $U_{\rm e}$, $U_{\rm h}$ or $U_{\rm x}$. The density distributions in Figs. 3b and 3a are from simulations with and without a perturbation via the electron potential $U_{\rm e}$, respectively. The amplitude of the disorder potential used, $A_{\rm rand}=1\,{\rm \mu eV}$ is orders of magnitude smaller than the fluctuations in $U_{\rm e}$ along the external line and the correlation length of the disorder ($2{\rm \mu m}$) is a few times less than the period of density modulation. This confirms that the density modulation is not correlated to the disorder but disorder is required to evolve the state of the line from the metastable solution shown in Fig\,3a to the stable solution shown in Fig\,3b. In the circular geometry, the numerical discretization of Eqs. (1-2) onto a rectangular grid provides some implicit distortion that triggers the fragmentation.

Fig. 3c shows the exciton density at the line position with and without disorder. A consequence of fragmentation is the reduction in the exciton density and, therefore, a lowering of the energy density associated with the dipole repulsion between excitons. We can conclude that in the simulations, the external line buckles into a wavy line in order to redistribute excitons over a larger area and reduce the energy density. We found that any non-zero perturbation triggers an energy minimization process, leading to fragmentation of the line.

The model presented provides insight into the role of Coulomb interactions in the formation of the external ring in the indirect exciton PL pattern. We have demonstrated that within a classical framework, periodic modulation of the exciton density on macroscopic length scales can occur. In the classical picture, fragmentation is a process of redistributing charges to minimize the potential energy associated with dipole repulsion. We found strong similarities with the available experimental data. In particular, the calculated temperature dependence of the ring contrast is in strong agreement.

We appreciate valuable discussions with R. Zimmermann and L.\,V. Butov. Computational work was performed using the facilities of ARCCA, Cardiff University. Support by the EPSRC is gratefully acknowledged.
\newline \newline $^\dagger$Deceased 29 August 2010.

\end{document}